\documentclass[prd,amsmath,showpacs,preprintnumbers,%
               superscriptaddress,floatfix,twocolumn,letterpaper]{revtex4}
\usepackage{amsmath}
\usepackage{graphicx}
\newcommand{\nc}{\newcommand}
\nc{\tr}{{\rm{tr}}}
\nc{\retr}{{\rm{Re\, tr}}}
\nc{\dg}{\dagger}
\nc{\hm}{\hat{\mu}}
\nc{\hn}{\hat{\nu}}
\nc{\eq}[1]{~Eq.~(\ref{#1})}
\nc{\Fmunu}{F_{\mu\nu}}
\nc{\Dmu}{{\mathcal D}_\mu}
\nc{\Dnu}{{\mathcal D}_\nu}
\nc{\fundDmu}{D_\mu}
\nc{\fundDnu}{D_\nu}
\nc{\scD}{{\mathcal D}}

\nc{\FSq}{\Fmunu^2(x)}
\nc{\DmuFSq}{(\Dmu\Fmunu(x))^2}
\nc{\DnuFSq}{(\Dnu\Fmunu(x))^2}
\nc{\DmuSqFSq}{(\Dmu^2\Fmunu(x))^2}
\nc{\DnuSqFSq}{(\Dnu^2\Fmunu(x))^2}
\nc{\DmuSqFDnuSqF}{\Dmu^2\Fmunu(x)\Dnu^2\Fmunu(x)}
\nc{\DmuDnuFSq}{(\Dmu\Dnu\Fmunu(x))^2}
\nc{\FSqSq}{\Fmunu^4(x)}
\begin{document}

\preprint{ADP-07-19-T659}
\begin{abstract}
  A new over-improved stout-link smearing algorithm, designed to
  stabilise instanton-like objects, is presented. A method for
  quantifying the selection of the over-improvement parameter,
  $\epsilon$, is demonstrated. The new smearing algorithm is compared
  with the original stout-link smearing, and Symanzik improved
  smearing through calculations of the topological charge and
  visualisations of the topological charge density.
\end{abstract}

\title{Over-Improved Stout-Link Smearing} \author{Peter J.\ Moran}
\author{Derek B.\ Leinweber} \affiliation{Special Research Centre for
  the Subatomic Structure of Matter (CSSM), Department of Physics,
  University of Adelaide 5005, Australia} \date{Dec 14, 2007}
\maketitle
\setlength{\arraycolsep}{1pt}
\section{Introduction}
\label{sec:introduction}

Studies of long distance physics in Lattice QCD simulations often
require the suppression of short-range UV fluctuations. This is
normally achieved through the application of a smoothing algorithm.
The most common prescriptions are cooling~\cite{cool1,cool2,cool3},
APE~\cite{ape1,ape2}, and improved APE smearing~\cite{Bonnet:2001rc},
HYP smearing~\cite{hyp} and more recently, EXP or stout-link
smearing~\cite{Morningstar:2003gk} and LOG
smearing~\cite{Durr:2007cy}.  Filtering methods such as these are also
regularly used in calculations of physical observables to improve
overlap with low energy states.

All smoothing methods require the use of an approximation to the
continuum gluonic action
\begin{equation}
  S_g = \frac{1}{2} \int \! d^4 x \ \tr[ F_{\mu\nu} F_{\mu\nu} ] \,.
  \label{eqn:continuumaction}
\end{equation}
Because we are approximating space-time by a 4-D lattice, these
approximations will contain unavoidable discretisation errors. These
errors can have a negative effect on the topological objects present
in the gauge field being studied and are detrimental to the smoothing
process.

Successful attempts have been made to remove the low order
discretisation errors by combining Wilson loops of different
shapes~\cite{symanzik,deForcrand:1995qq,Bonnet:2001rc,BilsonThompson:2002jk}.
Unfortunately, as shown by Perez,
\emph{et~al.}~\cite{GarciaPerez:1993ki} and reiterated below, it is
still possible for the remaining errors to spoil instanton-like
objects in the field. They proposed over-improved cooling as a means
of taming these errors via the introduction of a new tunable parameter
$\epsilon$ into their action~\cite{GarciaPerez:1993ki}.

In Sec.~\ref{sec:smoothingalgorithms} we begin by presenting a summary
of the most common smoothing algorithms. We then motivate the
introduction of over-improvement by considering the effect of the
discretisation errors on the action of a single instanton in
Sec.~\ref{sec:discretisationerrorsandimprovement}. Extending the work
of Perez, \emph{et~al.}~\cite{GarciaPerez:1993ki} we define in
Sec.~\ref{sec:overimprovement} an over-improved stout-link smearing
algorithm and demonstrate how to quantitatively select a value for the
parameter, $\epsilon$. Lastly, in Sec.~\ref{sec:algorithmcomparisons}
with the calibration of the over-improved parameter complete we test
the over-improved stout-link smearing on a variety of lattices,
including large $28^3\times96$ light dynamical gauge fields from the
MILC collaboration~\cite{Bernard:2001av,Aubin:2004wf}.

\section{Smoothing Algorithms}
\label{sec:smoothingalgorithms}

Standard cooling proceeds via a systematic sequential update of all
links $U_\mu(x)$ on the lattice, where at each link update the local
Wilson action~\cite{Wilson} is minimised. The local Wilson action
corresponding to $U_\mu(x)$ is defined as
\begin{equation}
  S_{W}(x) = \beta \sum_{\scriptstyle \nu \atop \scriptstyle \nu \ne \mu}
  \frac{1}{3} {\retr} \left [ 1 - U_{\mu}(x) \Sigma_{\mu \nu}(x) \right ] \, ,
  \label{eqn:localwilsonaction}
\end{equation}
where
\begin{equation}\label{eqn:wilsonstaples}
  \begin{split}
    \Sigma_{\mu \nu}(x) &= U_{\nu}(x+\hm) \, U^{\dg}_{\mu}(x+\hn) \, U^\dg_{\nu}(x)\\
    &\quad +U^{\dg}_{\nu}(x+\hm-\hn) \, U^{\dg}_{\mu}(x-\hn) \,
    U_{\nu}(x-\hn) \,
  \end{split}
\end{equation}
is the sum of the two staples touching $U_\mu(x)$ which reside in the
$\mu$-$\nu$ plane. From\eq{eqn:localwilsonaction}, we can see that
$S_W$ will be minimised when ${\retr} \left [ 1 - U_{\mu}(x)
  \Sigma_{\mu \nu}(x) \right ] = 0$.  It naturally follows that when
cooling, the aim is to replace $U_\mu(x)$ with a new link that
optimises
\begin{equation}
  {\rm{max}}\,{\retr} \left ( U_{\mu}(x) \sum_{\scriptstyle \nu \atop
      \scriptstyle\nu \ne \mu} \Sigma_{\mu \nu}(x) \right ) \, . 
  \label{eqn:coolingmax}
\end{equation}
When performing this update in parallel, one must be careful not to
replace any link which is included in the local action of a
neighbouring link.  This requirement means that cooling is a
relatively slow operation computationally, but fast in regard to the
removal of action from the gauge field.

APE smearing differs from standard cooling in that all links can be
simultaneously updated in a single sweep through the lattice,
resulting in a significant speed increase. In APE smearing one first
calculates a smeared link $U_\mu^\prime(x)$, which is the weighted sum
of its nearest neighbours,
\begin{equation}
  U_{\mu}^\prime(x) = (1-\alpha) \, U_{\mu}(x) + 
  \frac{\alpha}{6} \sum_{\scriptstyle \nu \atop \scriptstyle
    \nu \ne \mu} \Sigma_{\mu \nu}^\dagger(x) \, ,
  \label{eqn:apesmearing}
\end{equation}
where $\Sigma_{\mu \nu}$ is defined as in\eq{eqn:wilsonstaples}, and
$\alpha$ is a real parameter, usually set to $\approx 0.7$.  The new
link $U_\mu^\prime(x)$ is then projected back into the $SU(3)$ group
via some projection operator $\cal P$,
\begin{equation}
  \tilde{U}_\mu(x) = {\cal P} \, U_\mu^\prime(x) \, .
  \label{eqn:apeprojection}
\end{equation}
The projection of\eq{eqn:apeprojection} is necessary because we have
performed an additive step in\eq{eqn:apesmearing}, which is not an
$SU(3)$ group operation. The projection step is not uniquely defined,
but the preferred method is to select the new smeared link $U_\mu(x)$
such that it maximises
\begin{equation}
  \retr \left( U_\mu(x) \, U_\mu^{\prime \dagger}(x) \right) \, .
  \label{eqn:apeprojectioncondition}
\end{equation}
In the limit $\alpha \rightarrow 1$ we see that\eq{eqn:apesmearing}
becomes
\begin{equation}
  U_{\mu}^\prime(x) \rightarrow \frac{1}{6} \sum_{\scriptstyle \nu \atop
    \scriptstyle \nu \ne \mu} \Sigma_{\mu \nu}^\dagger(x) \, .
\end{equation}
Substituting this result into\eq{eqn:apeprojectioncondition} shows how
the projection method has become equivalent to
cooling\eq{eqn:coolingmax}, and that there exists a direct link
between APE smearing and cooling in the limit that links are updated
sequentially. The simultaneous update of APE smearing limits $\alpha <
0.75$~\cite{Bernard:1999kc}.

The more recent smearing technique, stout-link
smearing~\cite{Morningstar:2003gk}, makes use of the exponential
function to remain with the gauge group and remove the need for a
projection step. Beginning with the staples\eq{eqn:wilsonstaples},
define
\begin{equation}
  C_\mu(x) = \sum_{\scriptstyle \nu \atop \scriptstyle \nu \ne \mu}
  \rho_{\mu\nu} \Sigma_{\mu\nu}^\dagger(x) \, ,
  \label{eqn:stoutoldC}
\end{equation}
where we will choose an isotropic four-dimensional constant
$\rho_{\mu\nu} = \rho_{sm}$, but other selections are possible.  The
matrix $Q_\mu(x)$ defined by
\begin{align}
  Q_\mu(x) & = \frac{i}{2}(\Omega_\mu^\dagger(x) - \Omega_\mu(x)) \notag \\
  & \quad - \frac{i}{6}\tr(\Omega_\mu^\dagger(x) - \Omega_\mu(x)) \,,
\end{align}
with
\begin{equation}
  \Omega_\mu(x) = C_\mu(x) \, U_\mu^\dagger(x) \,,  
\end{equation}
is by definition Hermitian and traceless, and hence $e^{i Q_\mu(x)}
\in SU(3)$. The new smeared link is then defined by
\begin{equation}
  \tilde{U}_\mu(x) = \mathrm{exp}(i Q_\mu(x) ) \, U_\mu(x) \, .
  \label{eqn:stoutlinksmearedlink}
\end{equation}
An expansion of the exponential in\eq{eqn:stoutlinksmearedlink}
results in the same sum of paths, to first order in $\rho_{sm}$, as
for APE smearing~\cite{Morningstar:2003gk}.  Given this, and the
already established link between APE smearing and cooling, it follows
that there exists a connection between cooling and stout-link
smearing.  Indeed, simulations of Lattice QCD show that for any given
gauge field, the structures revealed by the smoothing procedures are
remarkably similar.

\vspace{1cm}

\section{Discretisation Errors and Improvement}
\label{sec:discretisationerrorsandimprovement}

The corrosion of topological structures during the smoothing process
is a well known side-effect of both cooling and
smearing~\cite{deForcrand:1995qq,Bonnet:2001rc,BilsonThompson:2002jk}.
It is the unavoidable discretisation errors in the lattice action that
are the cause of this observed behaviour.  This obviously inhibits our
ability to study the topological excitations on the lattice with the
most local operators and it would be beneficial if it could be
prevented.

When a gauge field is smoothed, the topological structures within are
subjected to the effects of lattice discretisation errors. One such
topological excitation is the instanton. To understand how the errors
will alter instanton distributions, first consider the clover Wilson
action given by
\begin{equation}
  S_{W} = \beta \sum_x \sum_{\mu > \nu} \left(
    1 - P_{\mu\nu}(x) \right ) \, ,
\end{equation}
where $P_{\mu\nu}(x)$ denotes $1/3$ of the real trace of the clover
average of the four plaquettes touching the point x.

Following Ref.~\cite{GarciaPerez:1993ki}, $S_{W}$ can be expanded in
powers of the lattice spacing, $a$, giving:
\begin{widetext}
  \begin{equation}
    \begin{split}
      S_{W} &= a^4 \sum_x \sum_{\mu > \nu} \tr \bigg{[} \frac{1}{2}
      \FSq
      - \frac{a^2}{24} \left( \DmuFSq + \DnuFSq \right) \\
      &- \frac{a^4}{24} \left( g^2 \FSqSq - \frac{1}{30} \left(
          \DmuSqFSq + \DnuSqFSq \right) - \frac{1}{3} \DmuSqFDnuSqF
        + \frac{1}{4} \DmuDnuFSq \right) \bigg{]} \\
      &+ O(a^{10},g^4) \, ,
    \end{split}
    \label{eqn:discretewilsonaction}
  \end{equation}
\end{widetext}

\newpage

where $i g F_{\mu\nu} = [ D_\mu , D_\nu ]$, $D_\mu = \partial_\mu + i
g A_\mu$, and $\scD_\mu \phi =~[ D_\mu , \phi ]$, for arbitrary
$\phi$.

The goal is to substitute the instanton solution~\cite{Belavin:1975fg}
given by
\begin{equation}
  A_\mu(x) = \frac{x^2}{x^2 + \rho_{inst}^2} \left( \frac{i}{g} \right)
  \partial_\mu (S) \, S^{-1} \, ,
  \label{eqn:instantonsoln}
\end{equation}
where
\begin{equation}
  S \equiv \frac{x_4 \pm i \, \vec{x} \cdot \vec{\sigma} }{ \sqrt{x^2}} \, ,
\end{equation}
for instantons/anti-instantons with $\sigma$ the Pauli matrices, into
the expanded Wilson action\eq{eqn:discretewilsonaction}. This requires
the use of the lattice approximation $a^4 \sum_x \approx \int \!  d^4
x$.  Substituting the instanton solution\eq{eqn:instantonsoln}
into\eq{eqn:discretewilsonaction} and performing the integration then
yields,
\begin{equation}
  S^{inst}_W = \frac{8 \pi^2}{g^2} \left[ 1 - \frac{1}{5} \left( \frac{a}{\rho_{inst}} \right)^{\!\!2} - \frac{1}{70}\left( \frac{a}{\rho_{inst}} \right)^{\!\!4} \right] \, .
  \label{eqn:instwilsaction}
\end{equation}
Notice that the leading error term in\eq{eqn:instwilsaction} is
negative and depends upon the instanton size $\rho_{inst}$. When the
Wilson action is used in a smoothing algorithm these errors result in
an under-estimation of the action density. Additionally, by decreasing
$\rho_{inst}$ the action will be further reduced. The smoothing
algorithms, which are trying to decrease $S$, will therefore shrink
$\rho_{inst}$ in order to reduce the action. Repeated application of
the smoothing procedures will eventually lead to overwhelming
discretisation errors and cause instantons to ``fall through the
lattice'' and disappear.

Improved actions aim to fix the problem of discretisation errors by
including different sized Wilson loops in the calculation of the
action, $S$. By choosing the coefficients of the loop combinations
carefully it is possible to eliminate the leading order error terms.

The Symanzik improved action uses a linear combination of plaquette
and rectangular loops to eliminate the $O(a^2)$ errors.
\begin{equation}
  \begin{split}
    S_{S} & = \beta \sum_x \sum_{\mu > \nu} \bigg[
    \frac{5}{3} ( 1 - P_{\mu\nu}(x) ) \\
    &- \frac{1}{12} \big( ( 1 - R_{\mu\nu}(x) ) + ( 1 - R_{\nu\mu}(x)
    ) \big) \bigg] \,.
  \end{split}
  \label{eqn:symanzikaction}
\end{equation}
Analogous to $P_{\mu\nu}$, $R_{\mu\nu}$ and $R_{\nu\mu}$ denote the
different possible orientations of the rectangular loops.

This can be expanded in terms of $a$, and the instanton solution
substituted as above to find, in agreement
with~\cite{GarciaPerez:1993ki}, that
\begin{equation}
  S_{S}^{inst} = \frac{8 \pi^2}{g^2} \left[ 1 - \frac{17}{210}\left( \frac{a}{\rho_{inst}} \right)^{\!\!4} \right] \, .
\end{equation}
The $O(a^2)$ error term has been removed by design, but we see that
the $O(a^4)$ term is still negative. Therefore, this action will still
not preserve instantons.

\vspace{1cm}

\section{Over-Improvement}
\label{sec:overimprovement}

\subsection{Formalism}
\label{subsec:formalism}

In 1993, Perez, \emph{et al.}~\cite{GarciaPerez:1993ki} introduced the
notion of over-improved cooling, also known as $\epsilon$-cooling.
The essential idea was that instead of trying to use different loop
combinations to completely eliminate higher order error terms, they
would instead choose their coefficients such that the error terms
become positive.

Introducing the parameter $\epsilon$, they defined the following
action,
\begin{equation}
  \begin{split}
    S_{P}(\epsilon) = \beta \sum_x \sum_{\mu > \nu} & \bigg{[}
    \frac{4-\epsilon}{3} ( 1 - P_{\mu\nu}(x) ) \\
    &+ \frac{\epsilon-1}{48} ( 1 - W_{\mu\nu}(x) ) \bigg{]} \, ,
  \end{split}
  \label{eqn:perezaction}
\end{equation}
where $W_{\mu\nu}(x)$ denotes the clover average of the $2\times2$
squares (windows) touching the point $x$.  Note that
in\eq{eqn:perezaction}, $\epsilon$ has been introduced such that
$\epsilon = 1$ gives the standard Wilson action and $\epsilon = 0$
results in an $O(a^2)$ improved action.  Expanding\eq{eqn:perezaction}
in terms of $a$ and substituting the instanton
solution\eq{eqn:instantonsoln} gives
\begin{equation}
  S_{P}^{inst} =  \frac{8 \pi^2}{g^2} \left[ 1 
    - \frac{\epsilon}{5} \left(\frac{a}{\rho_{inst}}\right)^{\!\!2}
    + \frac{4-5\epsilon}{70} \left(\frac{a}{\rho_{inst}}\right)^{\!\!4} \right] \, ,
  \label{eqn:perezinstanton}
\end{equation}
where the $O(a^2)$ term is directly proportional to $-\epsilon$. Thus,
by making $\epsilon < 0$ the leading order discretisation errors
become positive, and the modified action should preserve instantons.

In the interests of preserving locality we choose to use the
traditional combination of plaquettes and rectangles as in the
Symanzik improved action in preference to the combination of the
$1\times1$ and $2\times2$ loops used in~\cite{GarciaPerez:1993ki}.  We
now introduce the parameter $\epsilon$ into the Symanzik improved
action\eq{eqn:symanzikaction}.  By requiring that $\epsilon=0$ gives
the $O(a^2)$ improved Symanzik action, and that $\epsilon=1$ gives the
standard Wilson action. This implies the following form for the
action,
\begin{equation}
  \begin{split}
    S(\epsilon) & = \beta \sum_x \sum_{\mu > \nu} \bigg[
    \frac{5-2\epsilon}{3} ( 1 - P_{\mu\nu}(x) ) \\
    &- \frac{1-\epsilon}{12} \big( ( 1 - R_{\mu\nu}(x) ) + ( 1 -
    R_{\nu\mu}(x) ) \big) \bigg] \,.
  \end{split}
  \label{eqn:overimpaction}
\end{equation}
Performing the usual expansion in $a$ gives:
\begin{widetext}
  \begin{equation}
    \begin{split}
      S(\epsilon) &= a^4 \sum_x \sum_{\mu > \nu} \tr \bigg{[}
      \frac{1}{2} \FSq - \frac{\epsilon \, a^2}{24} \left( \DmuFSq +
        \DnuFSq \right)
      + \frac{a^4}{24} \bigg( g^2 (1-2\epsilon) \FSqSq \\
      &+ \frac{5\epsilon - 4}{30} \left( \DmuSqFSq + \DnuSqFSq \right)
      + \frac{2\epsilon - 1}{3} \DmuSqFDnuSqF
      + \frac{1 - 2\epsilon}{4} \DmuDnuFSq \bigg) \bigg{]} \\
      &+ O(a^{10},g^4) \,.
    \end{split}
  \end{equation}
\end{widetext}
Into which we substitute the instanton solution to find that
\begin{equation}
  S^{inst}(\epsilon) = \frac{8 \pi^2}{g^2} \left[ 1 
    - \frac{\epsilon}{5} \left(\frac{a}{\rho_{inst}}\right)^{\!\!2}
    + \frac{14\epsilon-17}{210} \left(\frac{a}{\rho_{inst}}\right)^{\!\!4} \right] \, .
  \label{eqn:overimpinstaction}
\end{equation}
As in\eq{eqn:perezinstanton}, negative values of $\epsilon$ will
result in a positive leading error term, and should preserve
instantons.

We introduce the over-improvement parameter into the stout-link
smearing algorithm by modifying the link combinations used
in\eq{eqn:stoutoldC}. Whereas the original $C_\mu(x) =
\rho_{sm}\,\sum\{ 1\times1{\rm\ paths\ touching\ }U_\mu(x) \}$, the
modified stout-link $C_\mu(x)$ has the form
\begin{equation}
  \begin{split}
    \label{eq:stoutcmux}
    \mathcal{C}_\mu&(x) = \rho_{sm}\,\sum\bigg\{ \frac{5-2\epsilon}{3}
    ({\rm 1\times1\ paths\ touching\ }U_\mu(x) ) \\
    &-\frac{1-\epsilon}{12} ({\rm 1\times2+2\times1\ paths\ touching\
    }U_\mu(x) ) \bigg\} \,,
  \end{split}
\end{equation}
and the smearing parameter $\rho_{sm}$ is unchanged. Note that both
forward and backward horizontally orientated rectangles are included
in the $2\times1$ paths, such that $\Omega_\mu(x)$ resembles the local
action.

\subsection{Tuning}
\label{subsec:tuning}

Of course, this now begs the question: How negative should $\epsilon$
be in order to preserve instantons?  Perez, \emph{et al}. reported a
value of $\epsilon = -1$ to preserve instantons, and indeed it
does. However, just as positive values of $\epsilon$ can shrink
instantons, so too can negative values cause instantons to grow. Just
as small instantons can fall through the lattice, big instantons can
grow so large that they are destroyed by the smoothing
procedure~\cite{BilsonNahm}. Additionally, one does not want to
unnecessarily distort the instanton-like objects in the gauge field.
Care must therefore be taken not to choose a value of $\epsilon$ that
is too negative.

In order to quantify the selection of $\epsilon$ we consider the ratio
$S(\epsilon) / S_0$, where $S_0 = 8 \pi^2 / g^2$ is the single
instanton action. Ideally $S(\epsilon) / S_0$ should be equal to 1 for
all values of the instanton size, $\rho_{inst}$, as it is in the
continuum.

\begin{figure}
  \begin{center}
    \includegraphics[angle=90,width=0.4\textwidth]{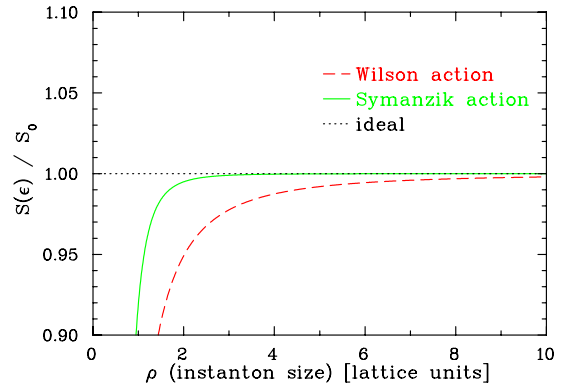}
  \end{center}
  \caption{$S(\epsilon) / S_0$ versus the instanton size for the
    Wilson and Symanzik improved actions. The ideal action would
    produce a flat line at $S(\epsilon) / S_0 = 1$. The positive slope
    on both curves means that instantons will shrink if the Wilson or
    Symanzik actions are used to smooth the gauge field.}
  \label{fig:wilsonsymanzik}
\end{figure}

Plots of $S(\epsilon) / S_0$ versus $\rho_{inst}$ for the Wilson and
Symanzik actions are shown in Fig.~\ref{fig:wilsonsymanzik}.  Note
that it is actually the slope of the curve that will govern whether an
instanton shrinks or grows, not just the sign of $S(\epsilon) / S_0$.
Although the Symanzik action is closer to the ideal action than the
standard Wilson action, the slope is still positive for all
$\rho_{inst}$ and using this action will shrink instantons.

\begin{figure}
  \begin{center}
    \includegraphics[angle=90,width=0.4\textwidth]{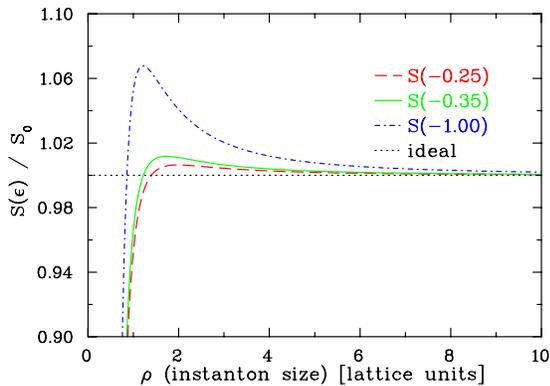}
  \end{center}
  \caption{$S(\epsilon)$ for three different values of $\epsilon$. The
    larger $-\epsilon$ is made the further the curve moves from the
    ideal behaviour and the sharper the maximum.}
  \label{fig:varyingepsilon}
\end{figure}

The goal is now to select a value of $\epsilon$ that results in the
flattest line possible, whilst ensuring the stability of instantons.
A plot for three different values of $\epsilon$ is shown in
Fig.~\ref{fig:varyingepsilon}.
With $\epsilon = -1$ the curve is similar to the mirror image of the
Wilson action. For $a \rho_{inst} > 1.5$, $\epsilon = -0.25$ and
$-0.35$ give curves closer to the ideal, however as $|\epsilon|$ is
decreased the maximum occurs at larger $\rho_{inst}$. Since it is the
slope that is responsible for how an instanton's size changes, the
maximum of $S(\epsilon)$ gives the dislocation threshold of the
smearing algorithm. Assuming that any topological excitation of length
$\ge 2a$ is not an unphysical UV fluctuation or a lattice artifact,
one should aim for a dislocation threshold of $\le 2a$.

Given this, we propose that a value of $\epsilon = -0.25$ will be
sufficient.  This choice gives a dislocation threshold of $\sim \!
1.97$, and a curve that is mostly flat down to values of $\rho_{inst}
\sim \!  1.7$. The action $S(\epsilon) / S_0$ is also very close to
the ideal.

\begin{figure}
  \begin{center}
    \includegraphics[angle=90,width=0.4\textwidth]{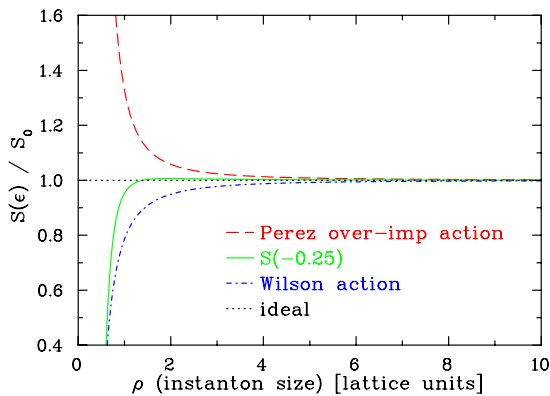}
  \end{center}
  \caption{A comparison of $S(\epsilon) / S_0$ for the Perez
    over-improved action, our over-improved stout-link smearing
    $S(-0.25)$, and the standard Wilson action.}
  \label{fig:perezourswilson}
\end{figure}

In Fig.~\ref{fig:perezourswilson}
we provide a comparison of the Perez over-improved action, our
over-improved action $S(-0.25)$, and the standard Wilson action.  It
is clear that $S(-0.25)$ should produce the best results.

Given a value for $\epsilon$ one can find a suitable value for the
smearing parameter, $\rho_{sm}$. Starting from some arbitrary value,
systematically increase $\rho_{sm}$ until $u_0$ (the mean-plaquette
value) no longer increases when smearing. This value sets an upper
threshold for $\rho_{sm}$ and one should then choose some $\rho_{sm}$
suitably below this threshold. In what follows we use a value of
$\rho_{sm} = 0.06$. A typical value for standard stout-link smearing
is $\rho_{sm}\approx 0.1$. The over-improved algorithm is more
sensitive to the smearing parameter than standard smearing because of
the larger loops used in the smoothing procedure.

\begin{figure}
  \begin{center}
    \includegraphics[angle=90,width=0.4\textwidth]{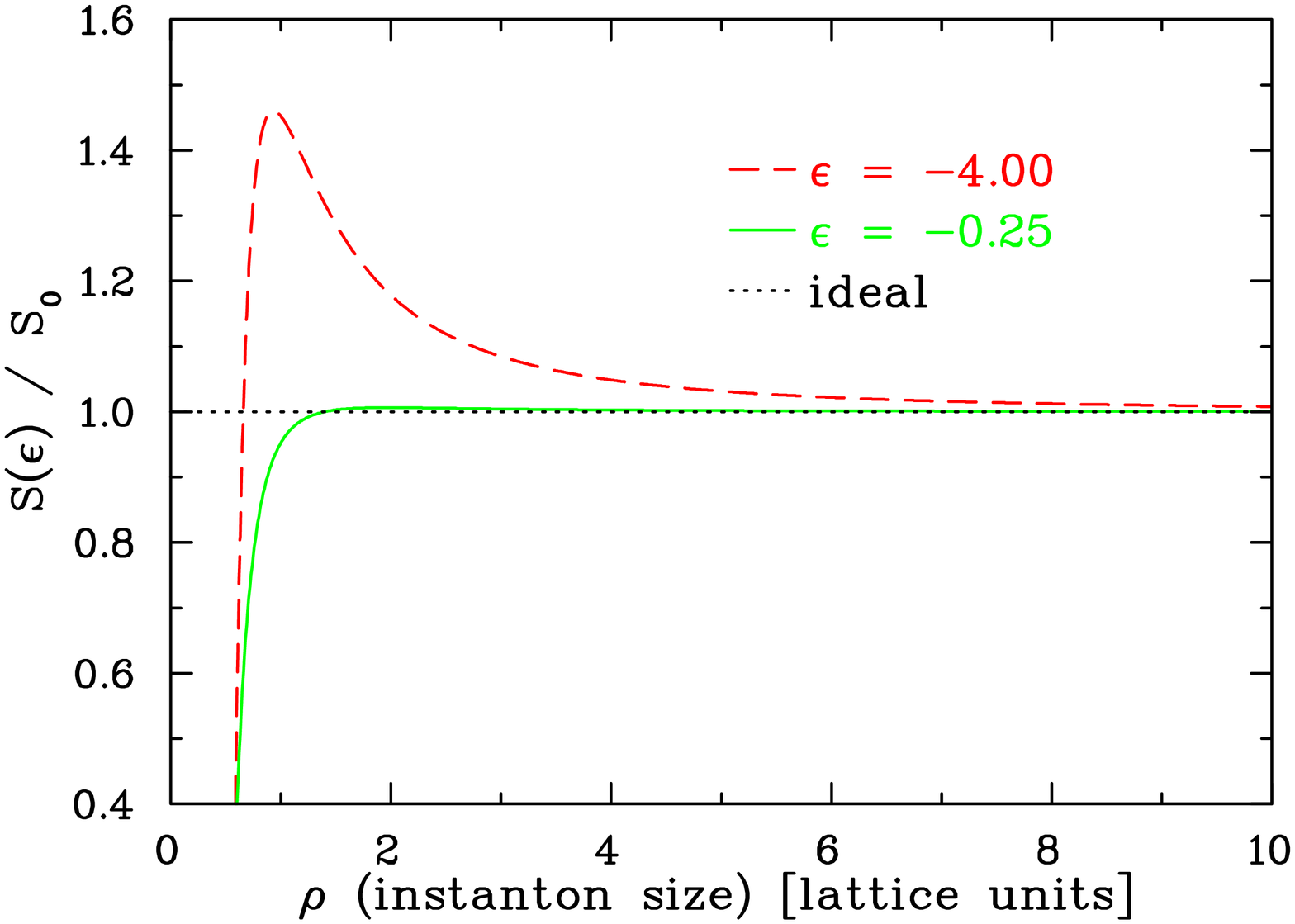}\\[8pt]
    \includegraphics[angle=90,width=0.38\textwidth]{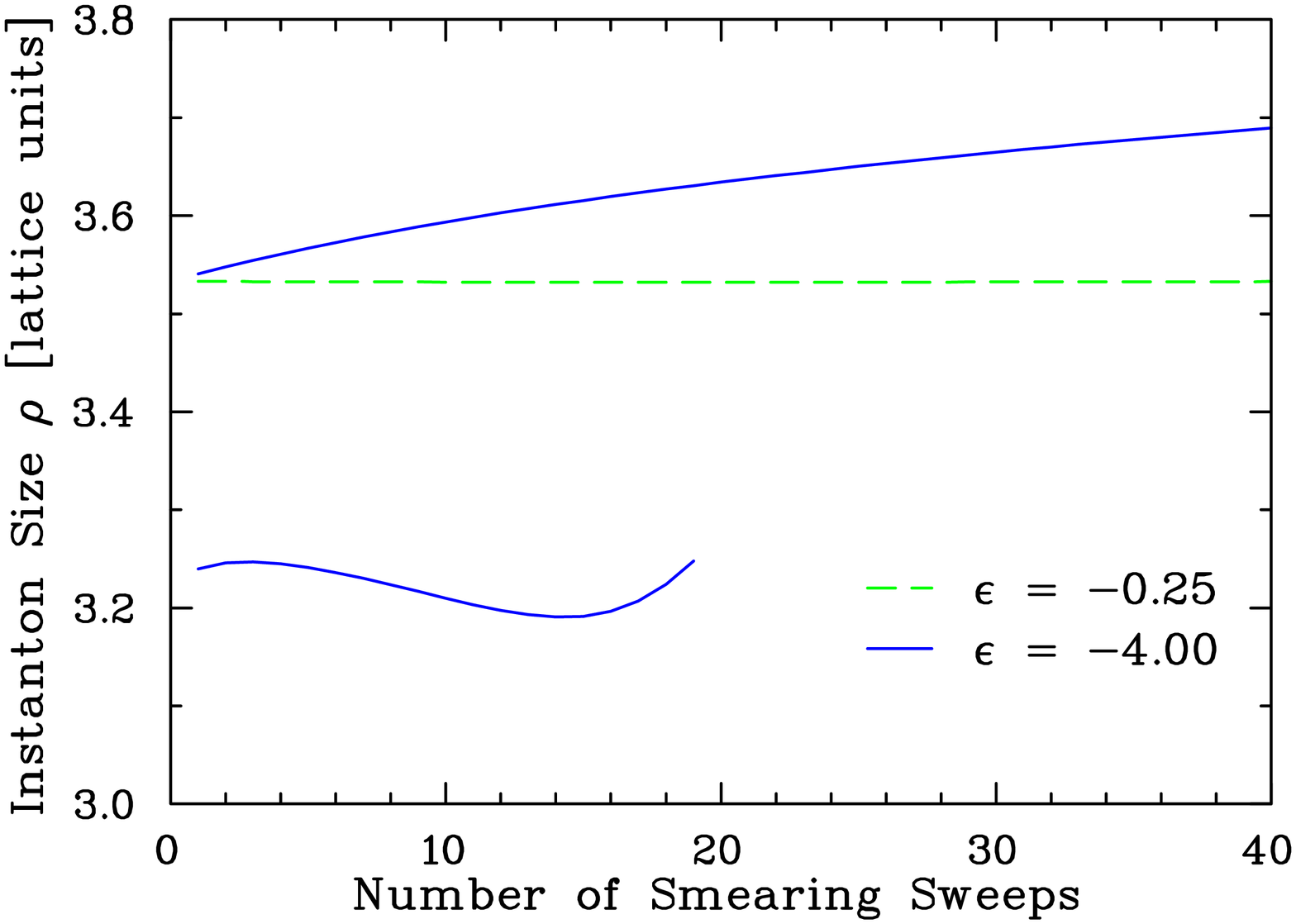}
  \end{center}
  \caption{The evolution of a single instanton under over-improved
    stout-link smearing with $\epsilon = -4.0$ and $\epsilon =
    -0.25$. The $S(\epsilon) / S_0$ plots for the respective
    $\epsilon$ values are shown in the top plot. The bottom plot shows
    how an instanton of size $\rho_{inst} \approx 3.5$ grows much more
    rapidly with $\epsilon = -4.0$ (upper solid line) than for
    $\epsilon = -0.25$ (dashed curve). The lower solid curve in the
    bottom plot is for an instanton of size $\rho_{inst} \approx 1$
    and $\epsilon = -4.0$.  It has been offset by $+1.8$ to fit on the
    same scale as the larger instanton.  We see how the instanton
    initially shrinks and is destroyed after 20 sweeps of
    over-improved stout-link smearing with $\epsilon = -4.0$.}
  \label{fig:singleinstantons}
\end{figure}

We mentioned earlier that it was the slope of $S(\epsilon) / S_0$ that
is responsible for how an instanton will evolve under smearing. We
thought it prudent to check that this was actually the case in
practice by smearing a single instanton gauge field. In order to
exaggerate the effects, we selected a rather extreme value of
$\epsilon = -4$. A comparison of $\epsilon = -4$ and $\epsilon =
-0.25$ is given in Fig.~\ref{fig:singleinstantons} .  Using $\epsilon
= -4$ to smear a single instanton of size $\rho_{inst} \approx 1$
should destroy the instanton. Meanwhile, if we smear an instanton with
size $\rho_{inst} \approx 3.5$ it should grow rapidly when $\epsilon =
-4$, and stay relatively the same size for $\epsilon = -0.25$.

Fig.~\ref{fig:singleinstantons} shows the results of the simulations.
The lowest curve is for the small instanton.  Note that we offset by
the size of the smaller instanton by $+1.8$ in order to display it on
the same scale as the larger instanton. As expected, the small
instanton initially shrinks under smearing until it is eventually
destroyed in the $20^{\rm th}$ sweep. For the larger instanton we see
that it does grow rapidly for $\epsilon = -4.0$, but that its size
stays relatively constant for $\epsilon = -0.25$.

\section{Algorithm Comparisons}
\label{sec:algorithmcomparisons}

Given the selection of $\epsilon = -0.25$ it is now important to make
a comparison of over-improved stout-link smearing with normal
stout-link smearing.  We are primarily concerned with the stability of
the topological charge under smearing, and the structure of the gluon
fields after smearing.

We use two sets of gauge fields for this study. Firstly, an ensemble
of large $28\times96$ dynamical MILC
lattices~\cite{Bernard:2001av,Aubin:2004wf}, with light quark masses;
$am_{u,d} = 0.0062$, $am_s = 0.031$.  We will also use a quenched MILC
ensemble of the same size and lattice spacing $a = 0.09$. The gauge
fields were generated using a Tadpole and Symanzik improved gauge
action with $1\times1 + 1\times2 + 1\times1\times1$ terms and an
Asqtad staggered dynamical fermionic action for the $2+1$ flavours of
dynamical quarks.

%
We also use some quenched CSSM gauge fields created with the
${\mathcal O}(a^2)$ mean-field improved L\"uscher-Weisz plaquette plus
rectangle gauge action \cite{Luscher:1984xn} using the plaquette
measure for the mean link.  The gauge-field parameters are defined by
\begin{equation}
  \begin{split}
    S_G = &\frac{5\beta}{3}\sum_{\scriptstyle x \atop \scriptstyle \nu
      > \mu}
    \left (1 - P_{\mu \nu}(x) \right) \\
    &-\frac{\beta}{12\, u_{0}^2}\sum_{\scriptstyle x \atop
      \scriptstyle \nu > \mu} \left (2 - R_{\mu \nu}(x) \right ) \, .
  \end{split}
\end{equation}
%
%
%
The plaquette measure of the tadpole improvement factor is
\begin{equation}
  u_0 = \left(\left \langle P_{\mu
        \nu}(x) \right \rangle_{x, \mu, \nu}
  \right)^{1/4} \, ,
  \label{uzero}
\end{equation}
where the angular brackets indicate averaging over space-time and
plaquette orientations.
%
%
%
The CSSM configurations are generated using the Cabibbo-Marinari
pseudo-heat-bath algorithm~\cite{Cabibbo:1982zn} using a parallel
algorithm with appropriate link partitioning \cite{Bonnet:2000db}.  To
improve the ergodicity of the Markov chain process, the three diagonal
SU(2) subgroups of SU(3) are looped over twice~\cite{Bonnet:2001rc}
and a parity transformation \cite{Leinweber:2003sj} is applied
randomly to each gauge field configuration saved during the Markov
chain process.

\subsection{Topological Charge}
\label{subsec:topologicalcharge}

Let us first consider the evolution of the total topological charge of
a gauge field under stout-link smearing. Typical studies in the past
have rated a smearing algorithm's success by its ability to generate
and maintain an integer charge. We will also use this test to rate the
effectiveness of the smearing procedures because of its simplicity and
widespread use.  However, it should be noted that we will be smoothing
extremely large $28^3\times96$ lattices. Due to the vast amount of
non-trivial topological charge field fluctuations present it will take
a lot of smoothing to generate an integer charge.

\begin{figure}
  \begin{center}
    \includegraphics[angle=90,width=0.31\textwidth]{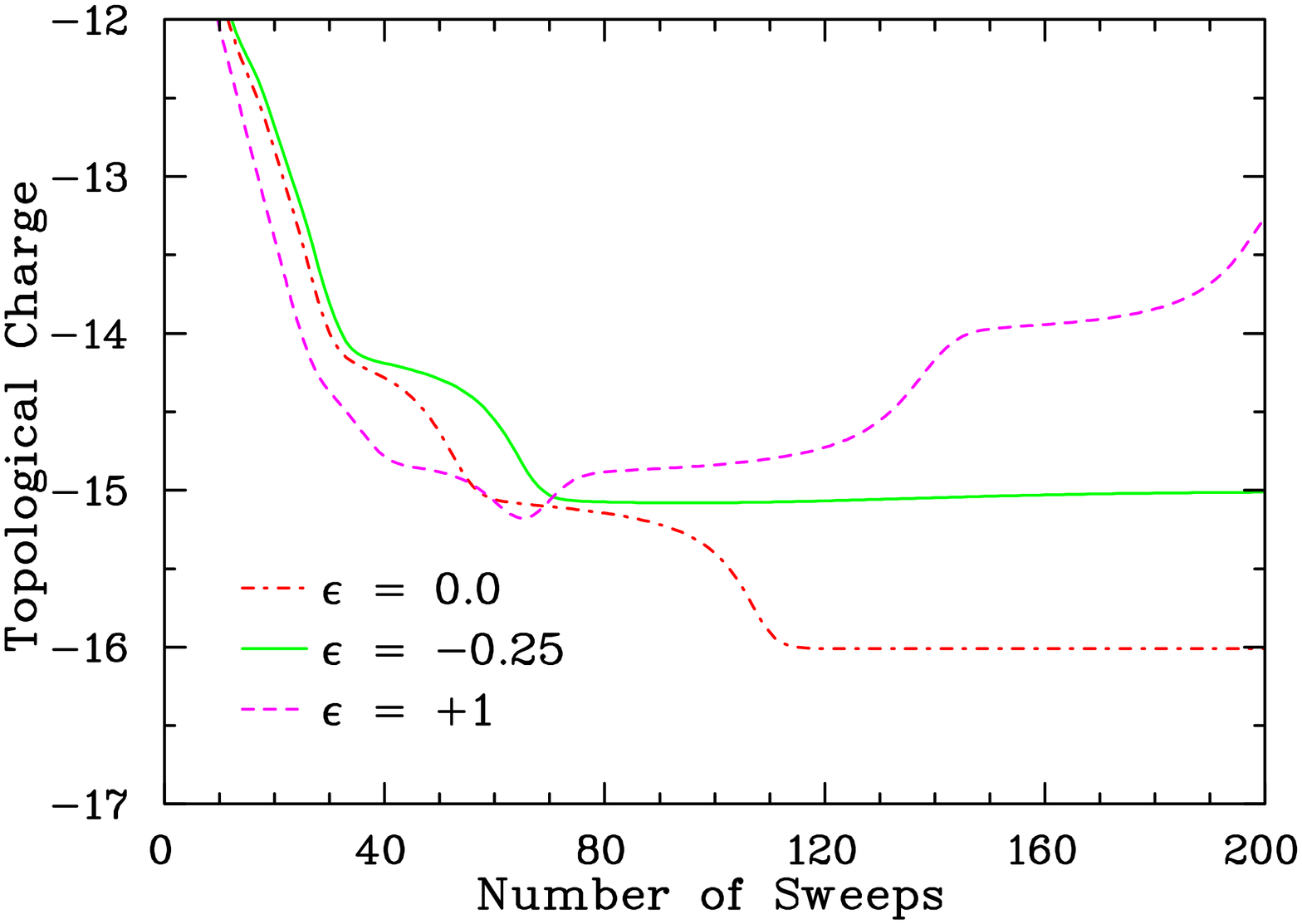}
    \\[5pt]
    \includegraphics[angle=90,width=0.31\textwidth]{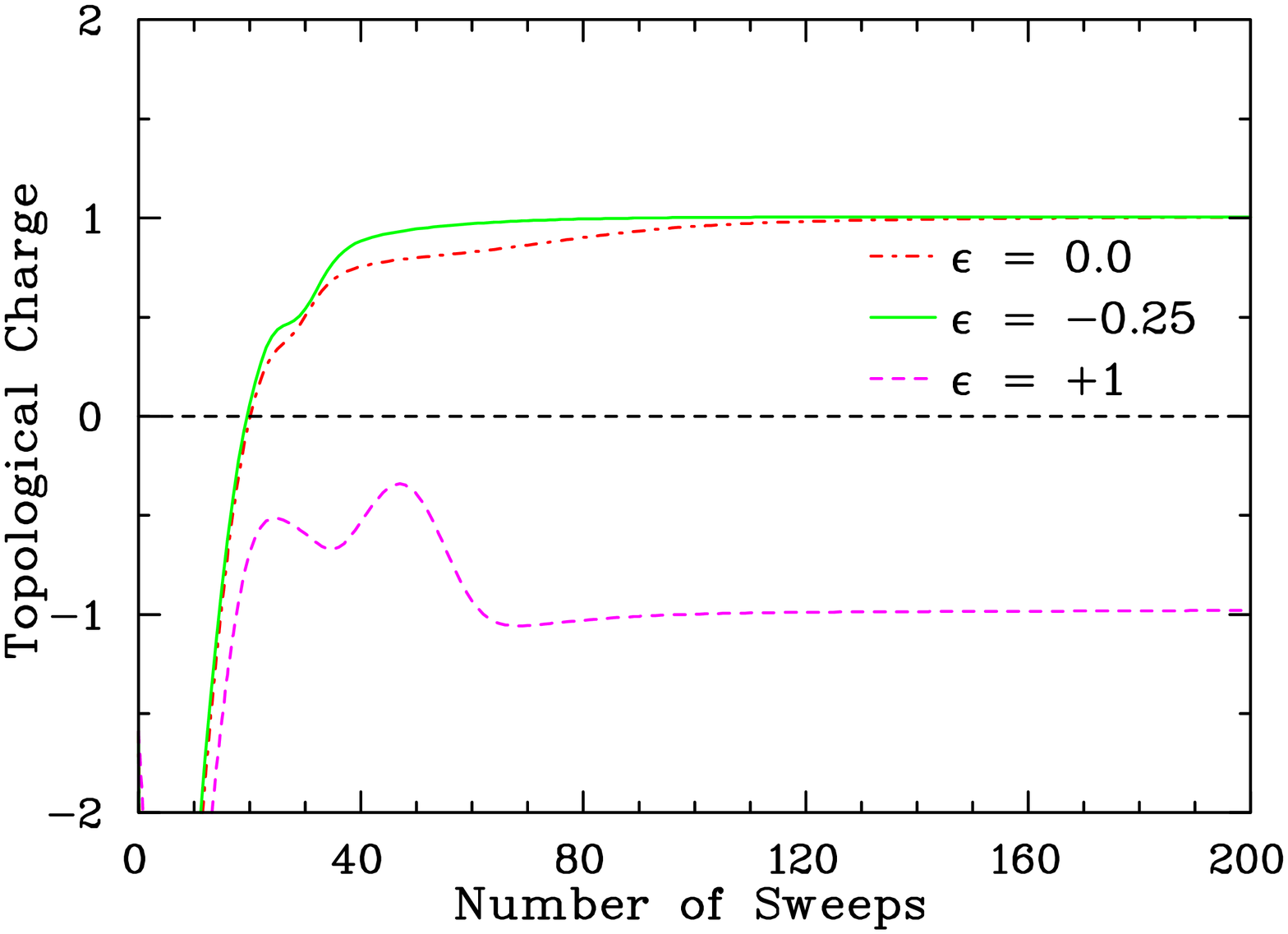}
    \\[5pt]
    \includegraphics[angle=90,width=0.31\textwidth]{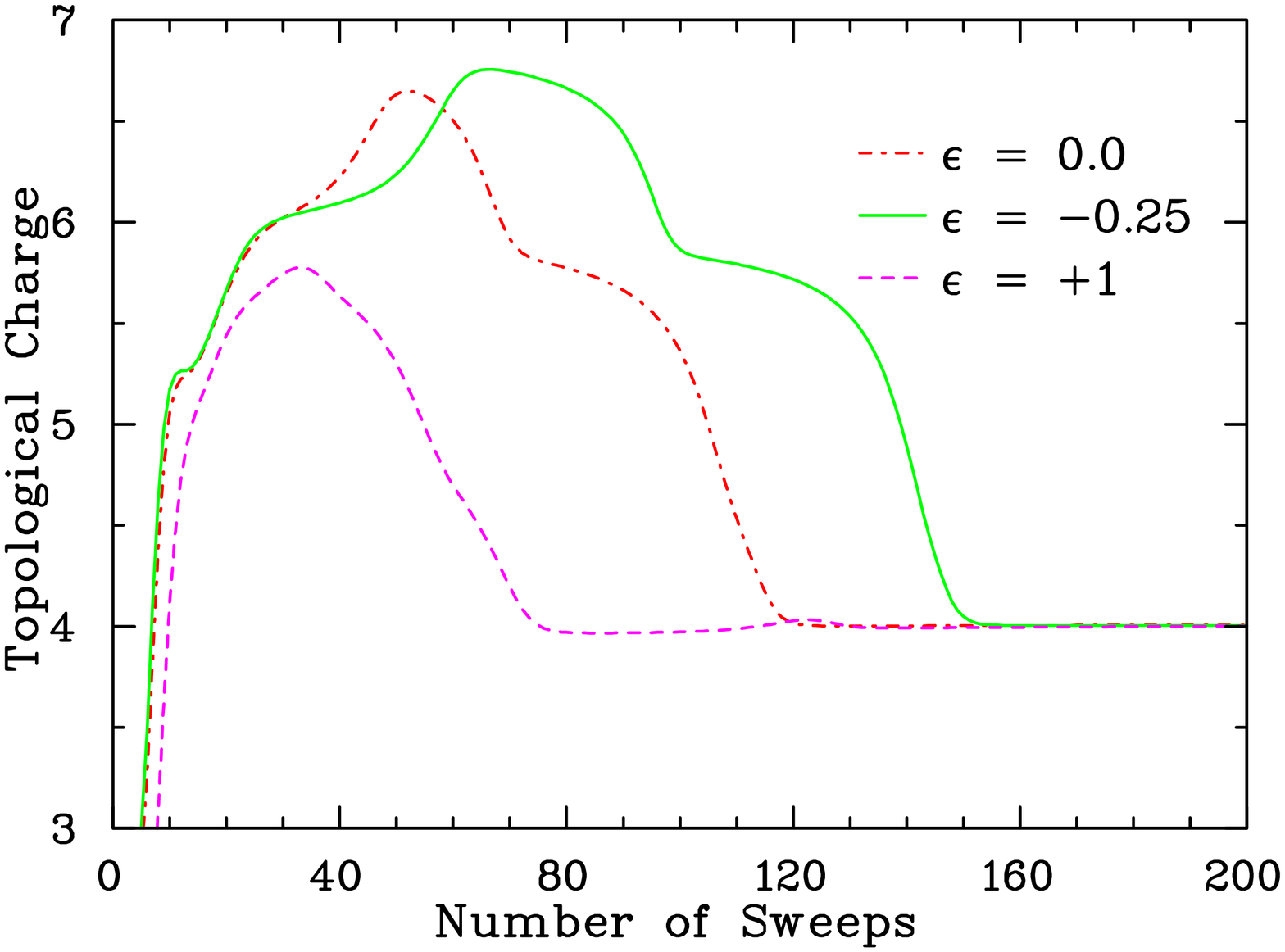}
    \\[5pt]
    \includegraphics[angle=90,width=0.31\textwidth]{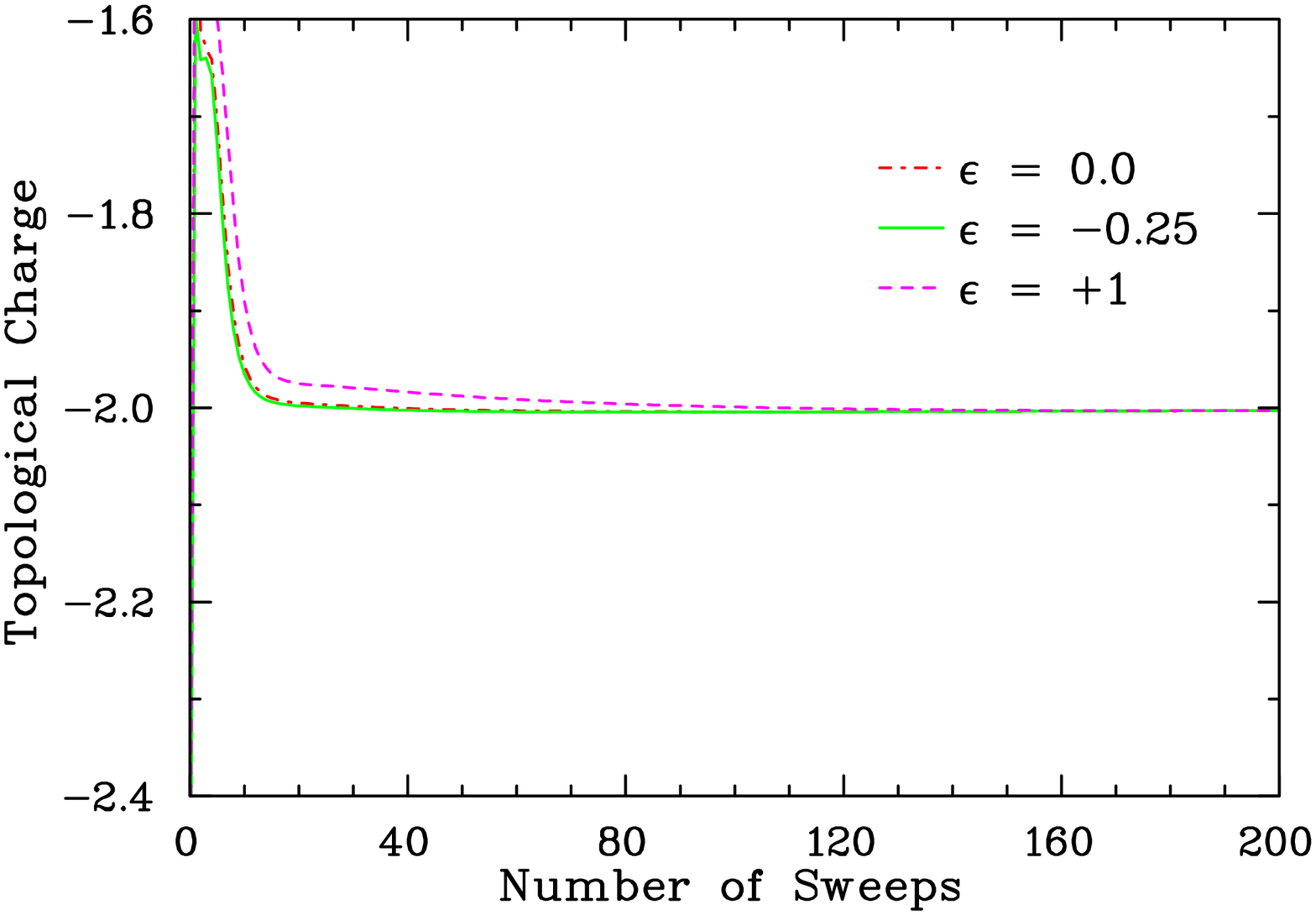}
  \end{center}
  \caption{Plots showing how the topological charge evolves under
    stout link smearing when using three different actions. The top
    graph is for a $28\times96$ quenched gauge field. The centre two
    are from an ensemble of $28\times96$ dynamical fields with light
    quark masses. The bottom is a smaller $16\times32$ quenched gauge
    field. The features of the graphs are explained in the main text.}
  \label{fig:topqevoln}
\end{figure}
Fig.~\ref{fig:topqevoln} provides a sample of 4 different gauge fields
smeared by standard, Symanzik improved, and over-improved stout-link
smearing. The first is a $28^3\times96$ quenched MILC gauge field, the
centre two are $28^3\times96$ light dynamical MILC fields, and the
last is a smaller $16^3\times32$ quenched field.

The top graph shows an example of the over-improved action producing a
stable result. In this instance the Wilson action is fluctuating
widely, and is unable to reach a stable charge within 200 sweeps of
smearing. The Symanzik improved action is better in that it stabilises
at around 120 sweeps, however the over-improved action is clearly
superior, stabilising 50 sweeps earlier.  At around 70-120 sweeps
there must exist a small instanton-like object that has been removed
by the errors in the Symanzik action, but preserved by the tuned
over-improved action.

The second graph is a more typical example of what one sees when using
the three different actions. The Wilson action is still clearly the
worst of the three, fluctuating the most. Meanwhile, the Symanzik and
over-improved actions are fairly similar in their behaviour. Both
stabilise at the same integer charge, but the over-improved action
stabilises earlier.

In the third graph we provide an example of how care must still be
taken when using over-improved stout-link smearing. Contrary to the
first two graphs, the curves in this graph seem to indicate that the
over-improved action is the worst of the three. In this gauge field
there must exist numerous instanton-like objects with size slightly
greater than 1. Objects with this size will still be removed by
over-improved smearing, but will survive for longer. Hence the
topological charge takes longer to stabilise. The effect was great in
this case because of the large lattice size, which meant it was
possible for a few of these objects to exist on the lattice. The
probability of finding such small objects on smaller lattices is
significantly less.

The final graph is a sample of a $16^3\times32$ lattice. It is shown
here to demonstrate how it is much easier to smooth a smaller gauge
field.  Note that not all small lattices are this simple to smooth and
we occasionally see behaviour similar to that in the top 3 graphs. In
the larger lattices, the larger size means that there is a greater
probability of finding an unstable topological object and it becomes
more difficult to achieve integer charges.

\subsection{Topological Charge Density}
\label{subsec:topologicalchargedensity}

For the next part of the analysis we will directly observe the
topological charge density of the gauge fields. Our aim is to directly
observe the differences in the gauge fields revealed by using the
Wilson and over-improved actions.

To achieve this we will require a gauge field where the final
topological charges from the two smearing procedures differ. We also
consider a smaller $16^3\times32$ lattice because smaller lattices
often provide clearer visualisations.

\begin{figure}
  \begin{center}
    \includegraphics[angle=90,height=0.2\textheight]{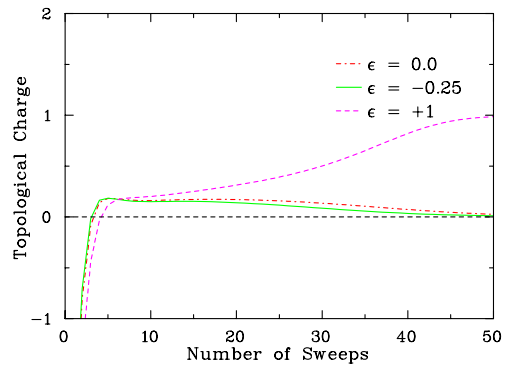}
  \end{center}
  \caption{The topological charge evolution under smearing for a
    $16\times32$ lattice. We see that when the Wilson action is used
    an anti-instanton is destroyed at around 20-40 sweeps. Some
    visualisations of the topological charge density in this region
    are provided later in the text.}
  \label{fig:qvsswpc003}
\end{figure}

The topological charge, as a function of the number of smearing
sweeps, is shown in Fig.~\ref{fig:qvsswpc003}.
It appears as though an anti-instanton is being destroyed by the
Wilson action from about $20$ sweeps onwards. It will be interesting
to visualise $q(x)$ in this region to see if we can observe this
behaviour.  Indeed, by considering the differences in the charge
density, we were able to locate the anti-instanton that is removed by
the Wilson action.

\begin{figure}
  \begin{center}
    \includegraphics[height=0.2\textheight]{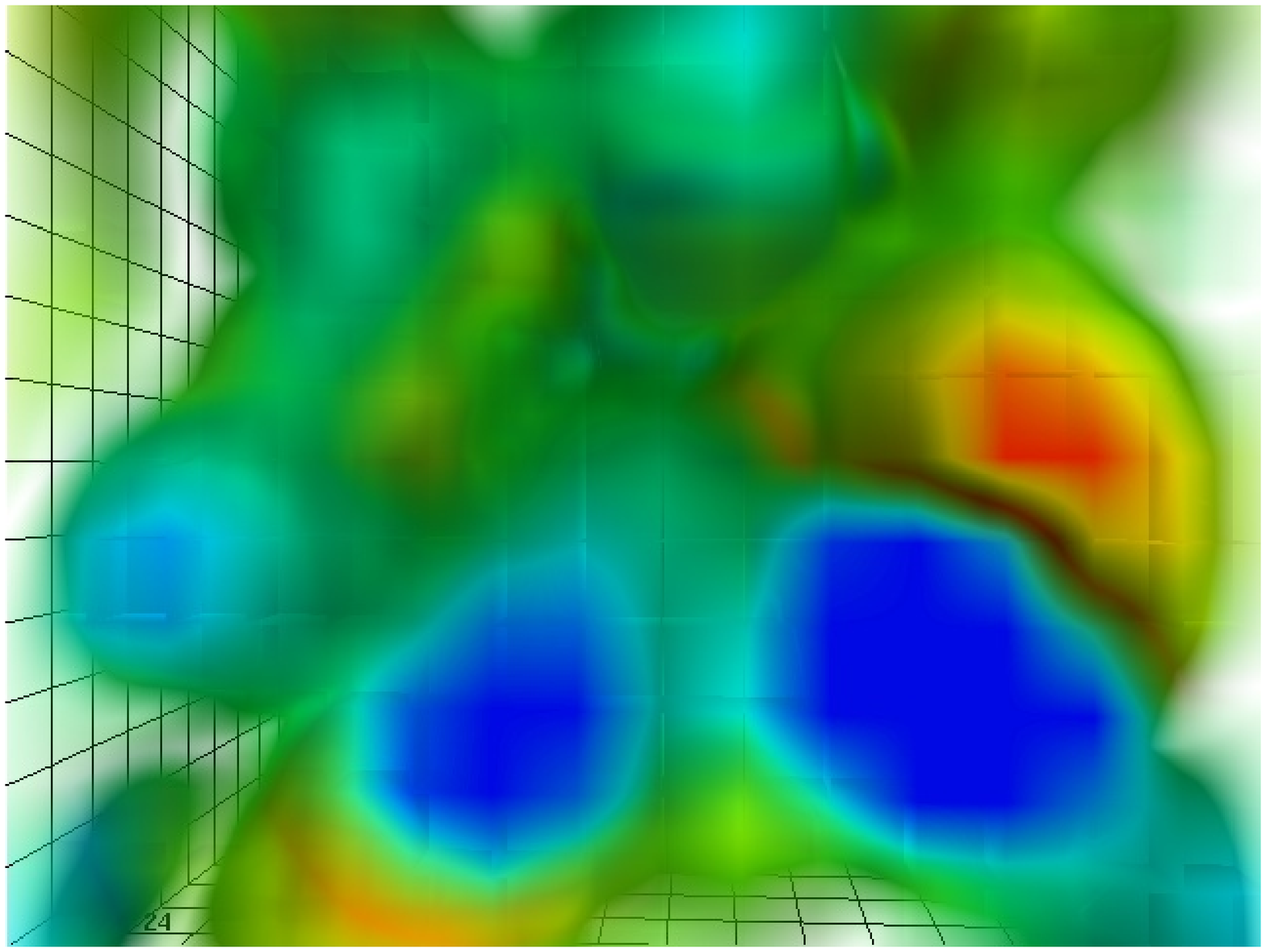}
    \includegraphics[height=0.2\textheight]{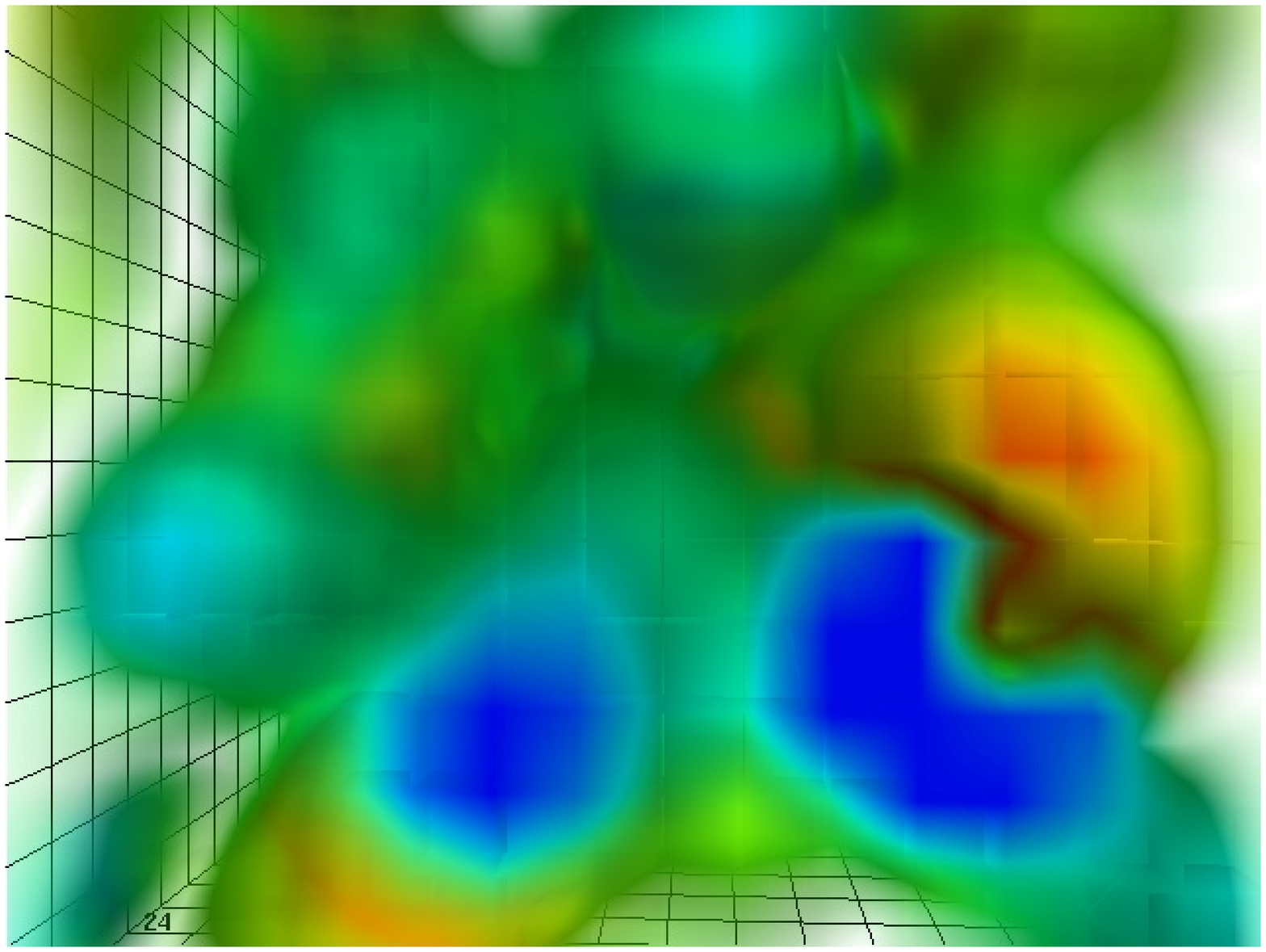}
    \includegraphics[height=0.2\textheight]{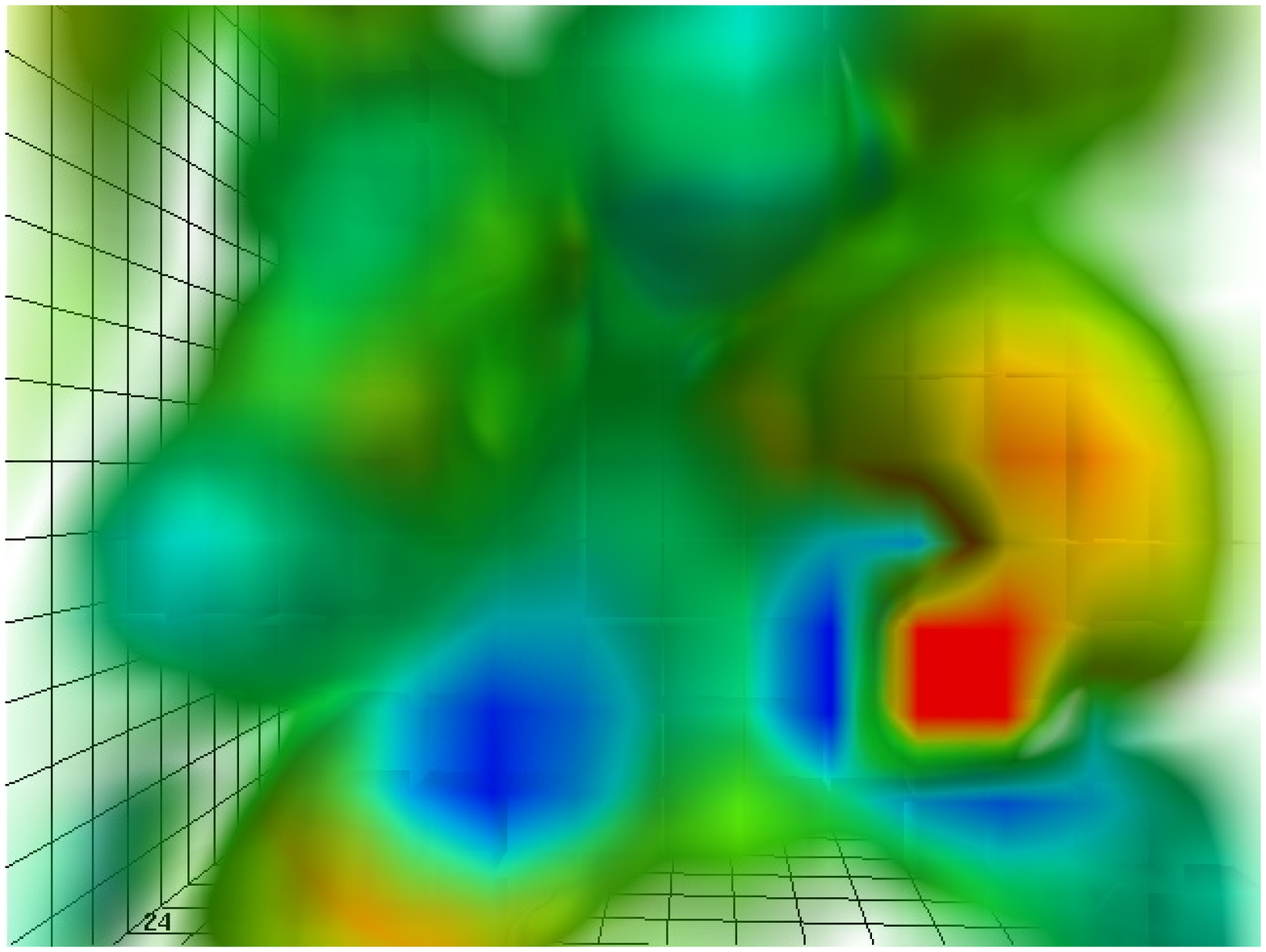}
    \includegraphics[height=0.2\textheight]{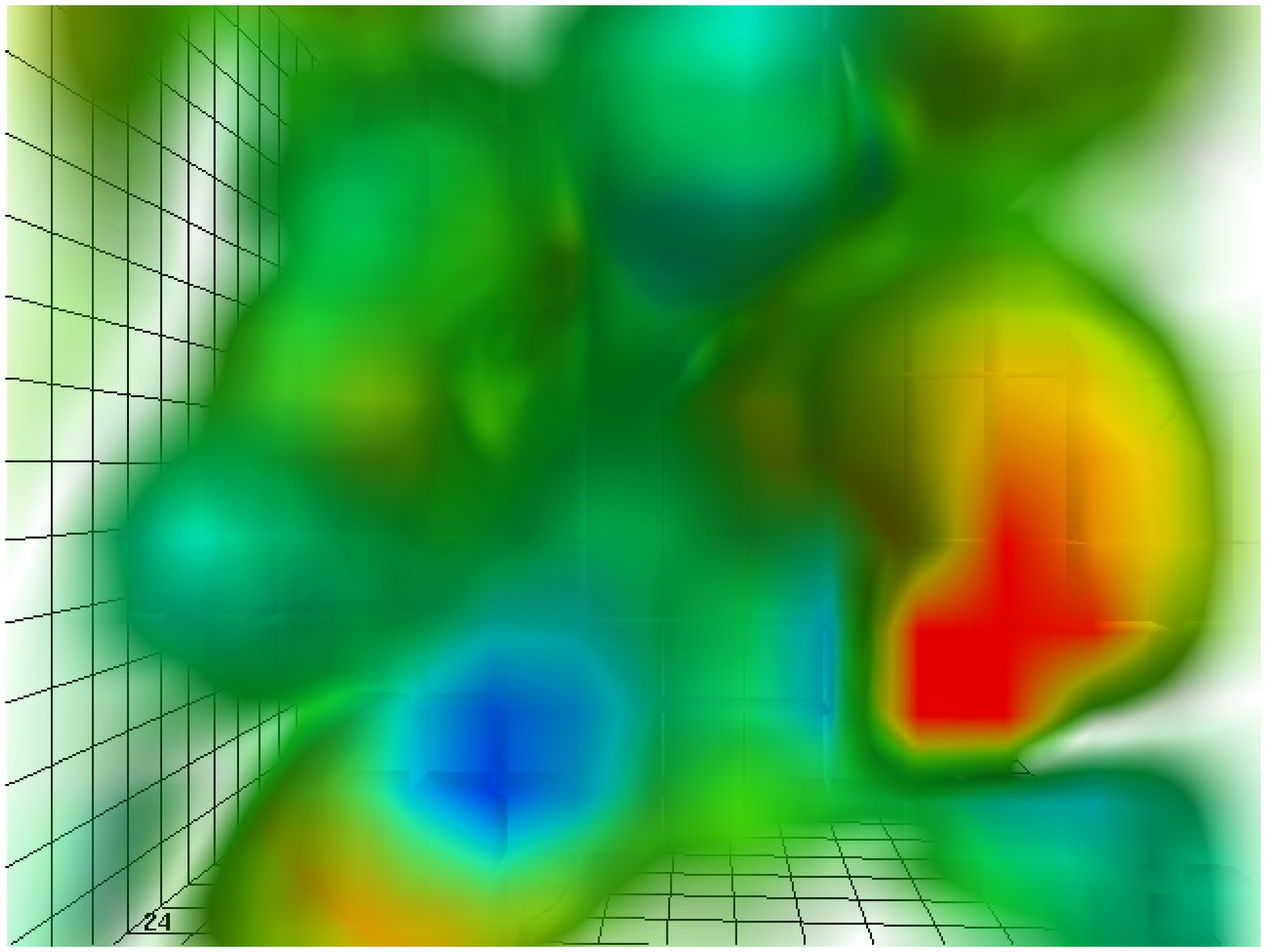}
  \end{center}
  \caption{The evolution of the topological charge density for various
    sweeps of standard stout-link smearing. The sweeps shown are; 30
    (top), 33, 36, 39 (bottom). In grey-scale (print) the darker areas
    represent regions of negative charge, and the lighter areas
    represent positive charge. In colour (online) blue to green
    represents negative topological charge and red to yellow
    represents positive. We see that a rather large anti-instanton is
    unstable under this smearing and is removed from the lattice,
    presenting an erroneous view of the vacuum.}
  \label{fig:evolnwilson}
\end{figure}

\begin{figure}
  \begin{center}
    \includegraphics[height=0.2\textheight]{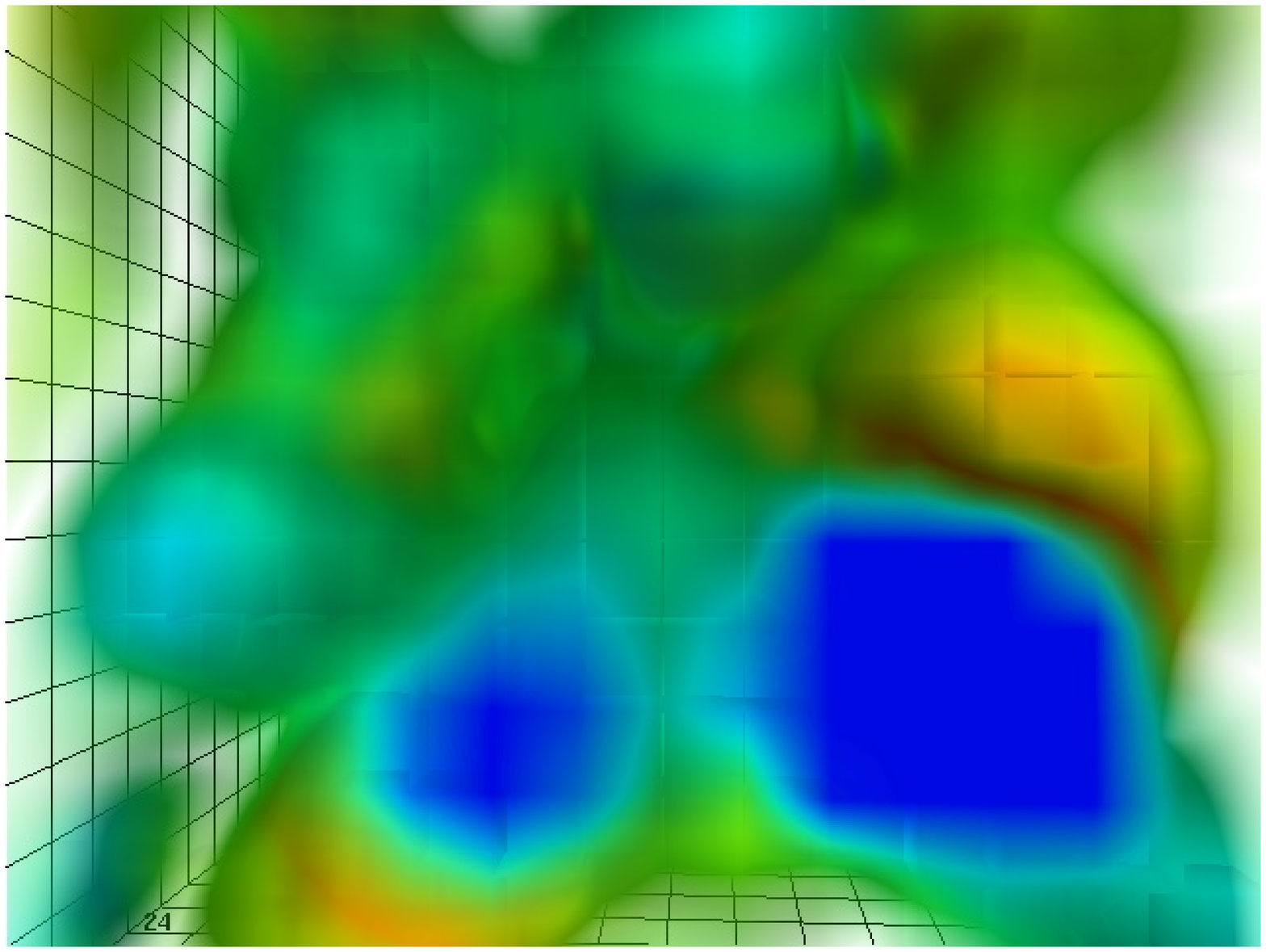}
    \includegraphics[height=0.2\textheight]{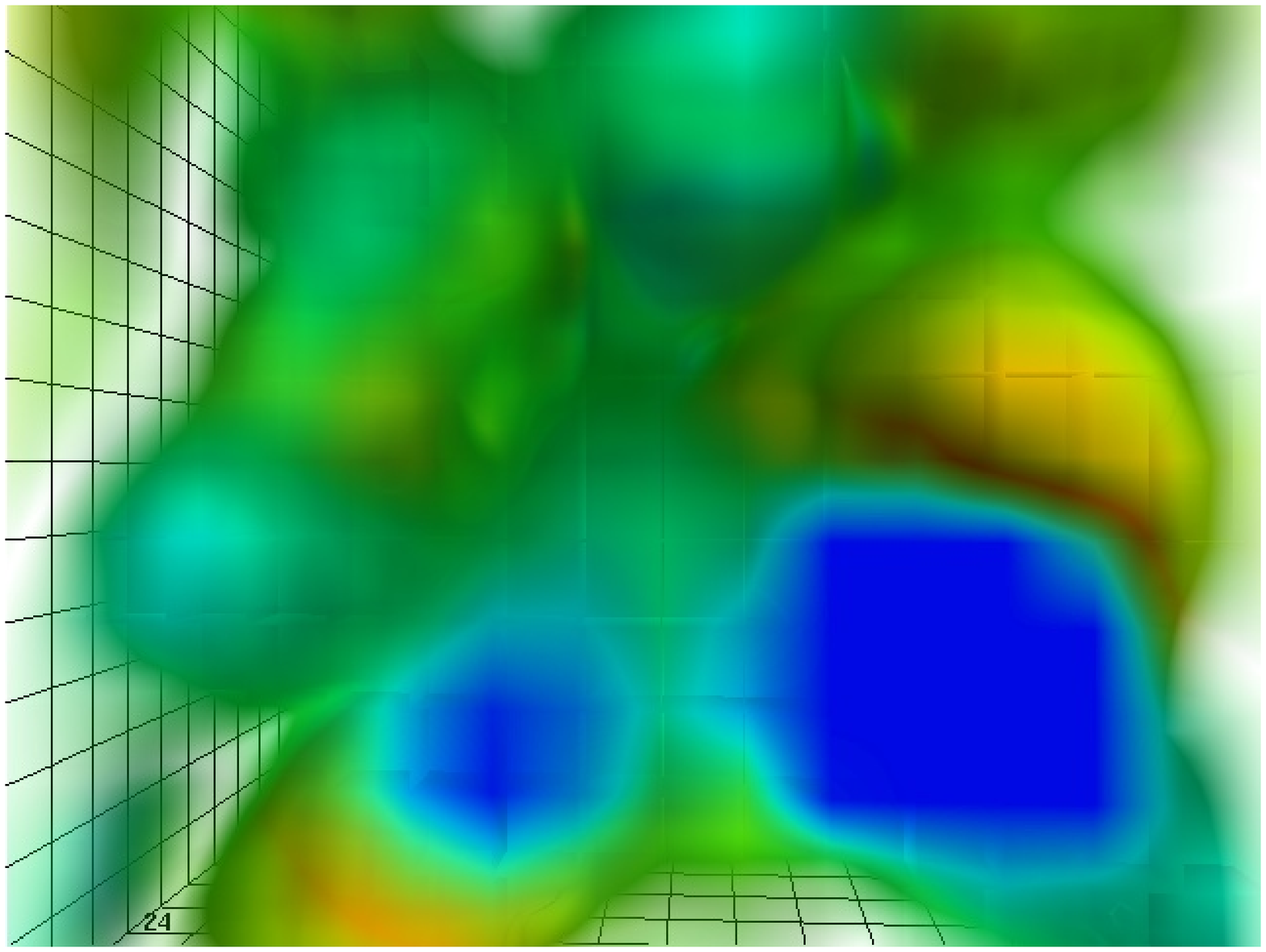}
    \includegraphics[height=0.2\textheight]{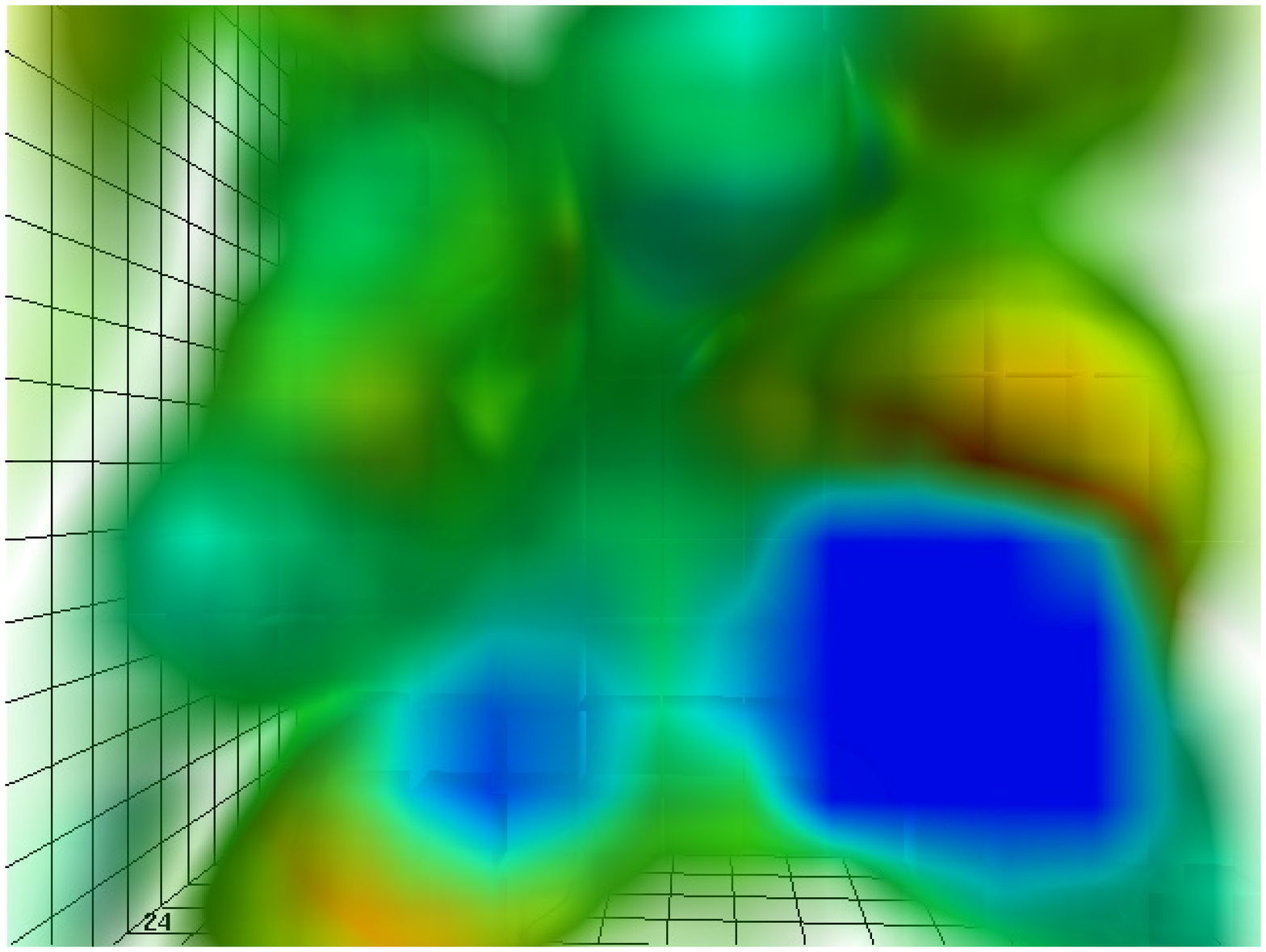}
    \includegraphics[height=0.2\textheight]{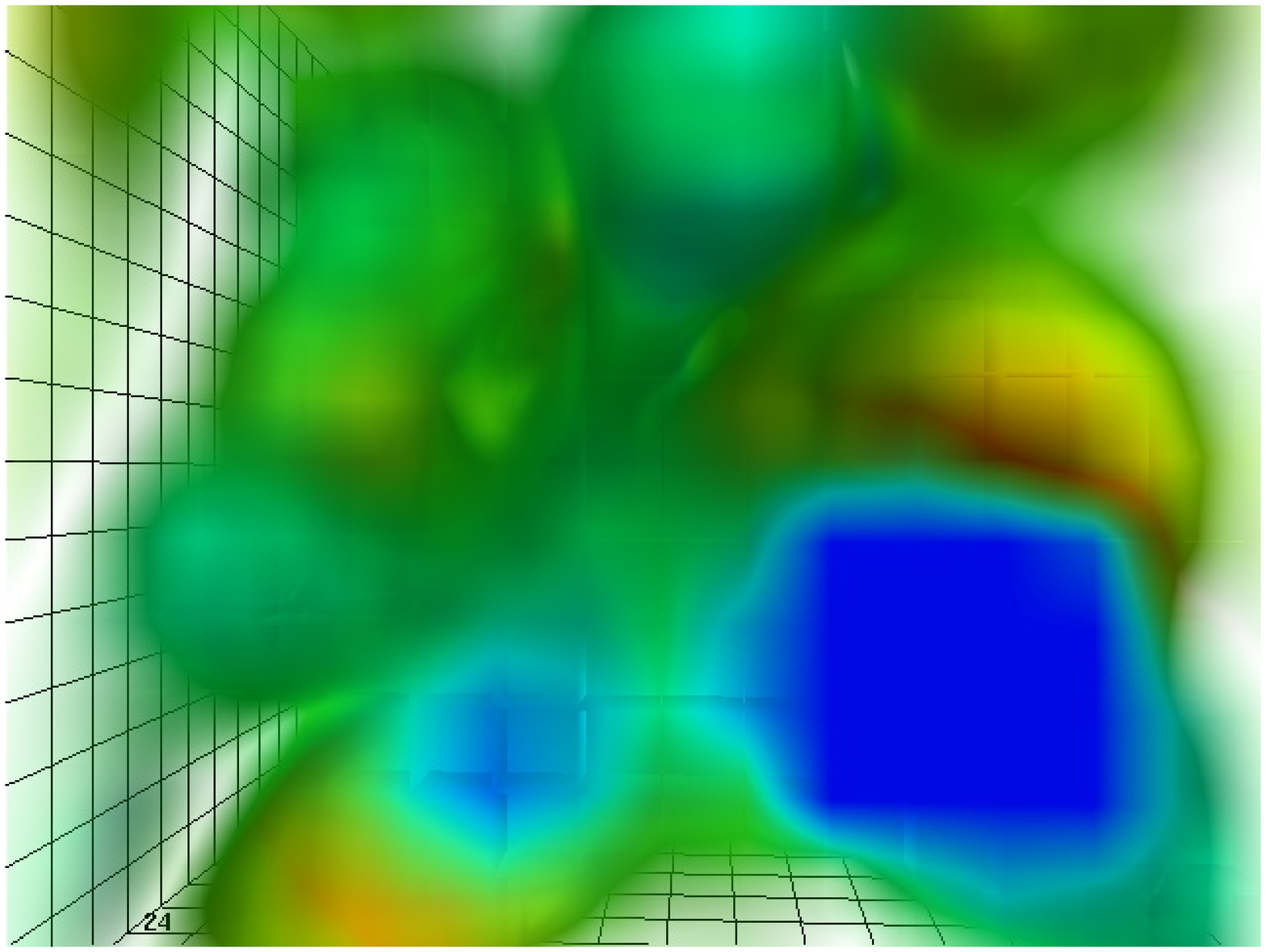}
  \end{center}
  \caption{A visualisation of the topological charge density of the
    same gauge field shown in Fig.~\ref{fig:evolnwilson}, this time
    with over-improved stout-link smearing. We see that in this case
    the anti-instanton in the lower right corner of the lattice is
    stable under smoothing, and remains stable for at least 200
    sweeps.}
  \label{fig:evolnoverimp}
\end{figure}

In Fig.~\ref{fig:evolnwilson} we show how the anti-instanton is
affected by the Wilson action, and in Fig.~\ref{fig:evolnoverimp} we
have the corresponding charge density from the over-improved action.
The pictures represent a single slice of the charge density of the 4-D
lattices as they evolve under the stout-link smearing. Objects with
negative topological charge are coloured blue to green, and the
positive objects are coloured red to yellow.

After 30 sweeps we see that both smearing methods have revealed a
similar vacuum structure. The effects of the errors in the Wilson
action are first seen after 33 sweeps, when the anti-instanton like
object on the right begins to unwind in the upper-right corner. Here
the charge density is approaching zero and therefore is not
rendered. In a few sweeps the action density in this region will
manifest itself in the opposite winding, largely eliminating the total
topological charge. The net effect is to suggest that the
instanton-like object on the right invades the neighbouring negative
object. However, the change in $Q$ indicates that this is not an
instanton - anti-instanton annihilation. At this point the majority of
the negative topological charge density is lost and the total $Q$ for
the configuration approaches $1$. This kind of phenomenon should not
be seen as filtering is applied to a lattice, and indeed it does not
occur when using the over-improved smearing.

After 36 sweeps the opposite winding has grown in size and it
continues to grow in size as more smearing is applied to the
lattice. After 39 sweeps the negatively charged object has all but
disappeared. Although not shown, eventually the neighbouring positive
object completely engulfs the region originally occupied by the
negatively charged excitation.

This is a direct demonstration of how the discretisation errors in the
Wilson action have resulted in an erroneous picture of the vacuum, and
how by modifying these errors in the over-improved algorithm we are
able to present a more accurate representation of the vacuum.

\section{Conclusion}
\label{sec:conclusion}

We have demonstrated how to define an over-improved stout-link
smearing algorithm, with the aim of preserving instanton-like objects
on the lattice. Using the new definition we showed how to select a
suitable value of the parameter $\epsilon$, and suggest a value of
$-0.25$.  With the procedure defined, we demonstrated the success of
the stout-link algorithm in preserving topological structures which
were destroyed when using the standard Wilson action.  The
over-improved stout-link smearing can be used in future studies of
vacuum structure or other similar applications, where preserving
topology on the lattice is important.

\section*{Acknowledgements}
\label{sec:acknowledgements}

The authors thank Waseem Kamleh for his assistance in interfacing with
his MPI Colour Orientated Linear Algebra (COLA) library.  We also
thank the Australian Partnership for Advanced Computing (APAC) and the
South Australian Partnership for Advanced Computing (SAPAC) for
generous grants of supercomputer time which have enabled this
project. This work is supported by the Australian Research Council.

\end{document}